# Electronic Origin of Half-metal to Semiconductor Transition and Colossal Magnetoresistance in Spinel HgCr$_2$Se$_4$

A. J. Liang[1,2,3*], Z. L. Li[4*], S. H. Zhang[1,2*], S. C. Sun[5], S. Liu[1], C. Chen[1,2,3,6], H. F. Yang[1], S. T. Cui[1], S.-K. Mo[3], S. Yang[4], Y. Q. Li[4], M. X. Wang[1,2], L. X. Yang[5], J. P. Liu[1,2†], Z. K. Liu[1,2†], Y. L. Chen[1,2,5,6†]

[1]*School of Physical Science and Technology, ShanghaiTech University, Shanghai, P. R. China*

[2]*ShanghaiTech Laboratory for Topological Physics, Shanghai 200031, P. R. China*

[3]*Advanced Light Source, Lawrence Berkeley National Laboratory, Berkeley, CA 94720, USA*

[4]*Beijing National Laboratory for Condensed Matter Physics,*
*Institute of Physics, Chinese Academy of Sciences, Beijing 100190, China*

[5]*State Key Laboratory of Low Dimensional Quantum Physics, Department of Physics and Collaborative Innovation Center of Quantum Matter, Tsinghua University, Beijing 100084, P. R. China*

[6]*Department of Physics, University of Oxford, Oxford, OX1 3PU, UK*

*These authors contributed equally to this work.

†Corresponding authors: liujp@shanghaitech.edu.cn，liuzhk@shanghaitech.edu.cn, yulin.chen@physics.ox.ac.uk

**Half-metals are ferromagnets hosting spin-polarized conducting carriers and crucial for spintronics applications. The chromium spinel HgCr$_2$Se$_4$ represents a unique type of half-metal, which features a half-metal to semiconductor transition (HMST) and exhibits colossal magnetoresistance (CMR) across the ferromagnetic-paramagnetic (FM-PM) transition. Using angle-resolved photoemission spectroscopy (ARPES), we find that the Fermi surface of *n*-type HgCr$_2$Se$_4$ (*n*-**

**HgCr$_2$Se$_4$) consists of a single electron pocket which moves above the Fermi level ($E_F$) upon the FM-PM transition, leading to the HMST. Such a Lifshitz transition manifests a giant band splitting which originates from the exchange interaction unveiled with a specific chemical nonstoichiometry. The exchange band splitting and the chemical nonstoichiometry are two key ingredients to the HMST and CMR, consistent with our *ab-initio* calculation. Our findings provide spectroscopic evidences of the electronic origin of the anomalous properties of HgCr$_2$Se$_4$, which address the unique phase transition in half-metals.**

The interplay of the spin, charge, orbital and lattice degrees of freedom in condensed matters provides important physical properties such as unconventional superconductivity [1], metal insulator transition [2], colossal magnetoresistance (CMR) [3] and anomalous Hall effect [4]. First proposed in the 1980's, half-metal is considered one such intriguing system [5]. A perfect half-metal hosts fully spin polarized conducting charge carriers that are ideal for spintronics applications. In the past decades, a handful of half-metals have been proposed, such as oxides [6-9] (*e.g.* CrO$_2$ [7]), spinels [9-10] (*e.g.* Fe$_3$O$_4$ [9]), perovskites [11-12] (*e.g.* (La,Sr)MnO$_3$ [3, 11]), Heuslers [5, 13-14] (*e.g.* NiMnSb [5]) and zinc blendes (*e.g.* CrAs [15]). Among them only few are experimentally verified to be half-metal with nearly 100% spin polarization [7, 11].

Recently, the Cr-based chalcogenide spinel HgCr$_2$Se$_4$ have been theoretically proposed [16-17] and experimentally verified as a half-metal [18]. It possesses magnetic moments of 3$\mu_B$/Cr$^{3+}$ with a spin polarization of the conducting carriers up to 97% [18]. Similar to mixed-valence manganites [3, 19-20] and Eu-based chalcogenides [21-23], HgCr$_2$Se$_4$ shows an intriguing resistivity anomaly across the ferromagnetic-paramagnetic (FM-PM) transition, where the resistivity changes by several orders in magnitude, featuring a half-metal to semiconductor transition (HMST) [18, 24-28]. The CMR [24-25, 27] is observed in the same temperature regime as HMST (~20K above and below T$_C$ [25]). Moreover, HgCr$_2$Se$_4$ also hosts spiral like antiferromagnetic insulating state under high pressure [29], $\sqrt{T}$-type dependence of versatile transport coefficients at low temperature [30], and may realize a Chern semimetal state [16] and other topological nontrivial states [17, 31-34]. These unique properties make $HgCr_2Se_4$ an important platform to investigate the interplay between the half-metallicity, magnetism and topology.

Despite the strong motivations, the electronic origin of the HMST, CMR and other intriguing properties [35-36] in $HgCr_2Se_4$ remain elusive. Further, the proposed topologically nontrivial band structure [16-17] requires experimental verification. Nonetheless, the photoemission spectroscopy investigation on the electronic structure of $HgCr_2Se_4$ has not been achieved. Hence, a direct measurement of the electronic band structure would offer crucial evidences in answering these questions.

In this work, using angle-resolved photoemission spectroscopy (ARPES), we report a comprehensive electronic structure study on $n$-$HgCr_2Se_4$. Far below $T_C$, we observed a Fermi surface produced by a single conduction band (CB) [17] and a characteristic ~0.3 eV direct band gap. Upon increasing temperature, the CB moves towards Fermi level ($E_F$) and eventually above it above $T_C$, showing a Lifshitz transition. Meanwhile the valence band (VB) remains almost unshifted, resulting in an increase of the band gap. Such an electronic evolution could be well reproduced by our *ab-initio* calculations, which suggests the shift of CB is driven by the exchange interaction that splits bands. The resulted CB is thus fully spin polarized below $T_C$. Our findings reveal that the Lifshitz transition which can be tuned by both the chemical nonstoichiometry ($E_F$ position) and the magnetic-field-mediated exchange band splitting are responsible for the ideal HMST (Fig. 1a) and CMR near $T_C$. These findings reveal the electronic origin of the most unusual properties in $HgCr_2Se_4$ and can be applied to related half-metals.

**Methods**

ARPES measurements were performed at BL 10.0.1, BL 7.0.2, BL 4.0.3 of the Advanced Light Source, BL 5-2 of the Stanford Synchrotron Radiation Light Source. BL 10.0.1 and BL 4.0.3 are equipped with R4000 analyzers. BL 7.0.2 and BL 5-2 are equipped with DA30 analyzers. The measured sample temperature and vacuum level were 20K~250 K and lower than $5.0\times10^{-11}$ Torr. The angle resolution was 0.2° and the overall energy resolution was better than 20 meV. Single crystals of *n*-HgCr$_2$Se$_4$ were grown by vapor transport method [30]. The Curie temperature $T_C$ is ~ 105 K with the electron density in a range of $10^{16}$–$10^{18}$ cm$^{-3}$ [30]. All the calculations are generated by PBE+U method, with a coulomb energy *U* on the Cr element of 5.5 eV.

**Results**

HgCr$_2$Se$_4$ crystallizes in the space group Fd3m with Cr atoms forming tetrahedrons (Fig. 1b). The increase of the lattice by Se makes the FM Cr-Se-Cr interaction prevail the Cr-Cr antiferromagnetic interaction [37], resulting in a FM transition at ~105K [18, 28, 30]. High quality samples (Fig. 1c(i)) have octahedral shape, flat shining surfaces and sharp edges with sharp Bragg peaks (Fig. 1c(ii)) and core levels (Fig. 1g). Fig. 1d shows the bulk and projected (111) surface Brillouin zones (BZs). The *n*-type nature (the as-grown sample is moderately *n*-doped [25]) is confirmed by Hall resistivity measurements [38-39]. The temperature dependent resistivity indicates a FM-PM transition at ~105 K with a resistivity variation by several orders in magnitude (Fig. 1e). Below $T_c$, the magnetic susceptibility shows clear soft ferromagnetism (Fig. 1f). Above $T_c$, the magnetic susceptibility curves gradually approach the linear shape, suggesting strong FM fluctuations above $T_c$.

The band structure of *n*-HgCr$_2$Se$_4$ in the FM state on (111) surface is illustrated in Figure 2. The freshly cleaved surface is slightly less *n*-doped comparing to the as-grown surface because of the rich interior Hg vacancies [25] (SI1). Therefore, the photoemission spectrum observed on the as-cleaved samples only contain hole-like bands (Figs. 2a-c). The constant energy contour (CEC) near the VB maximum shows a clear six-fold symmetry, consistent with the (111) surface cleavage (Fig. 2a). By using both the linear horizontal (LH) and vertical (LV) polarized photons, four sets of hole-like bands ($\alpha$, $\beta$, $\gamma$, $\delta$) along $\bar{\Gamma}$-$\bar{M}$ (Figs. 2b-c) are identified, agreeing well with the calculation (Fig. 2d).

To measure the bandstructure of the CB, we raise the $E_F$ by using *in situ* surface alkaline metal (K) deposition (SI2). Upon K dosing, the CB ($\varepsilon$ band, a small electron pocket at $\bar{\Gamma}$ as labelled in Figs. 2d) appears below $E_F$ (Figs. 2e-h). Along the $\bar{\Gamma}-\bar{M}$ direction, the CB and VB (uppermost $\alpha$ band) have effective masses of 0.15 $m_e$ and 2.88 $m_e$, and a ~0.3eV direct band gap at 20K (Fig. 2g). The CB and VB around $\bar{\Gamma}$ can also be observed by spectrum on the (001) cleaved surface (SI3). By extracting the band dispersion along $\Gamma-L$ (Fig. 2h), we further confirm there is no in-gap state. All the observed bands ($\alpha$, $\beta$, $\gamma$, $\delta$, $\varepsilon$) and the gap size are captured by our first-principle calculation (Fig. 2d). From the calculations we conclude that the $\alpha,\gamma$ and $\beta,\delta$ bands are spin-split pairs in the FM state. Since the calculated spin splitting of the CB is ~1 eV [16-17, 37], the observed $\varepsilon$ band is thus fully spin polarized (its spin split counterpart lies way above $E_F$), consistent with the half-metal nature in *n*-HgCr$_2$Se$_4$ [18].

We estimate the evolution of the CB, VB and the band gap across $Tc$ by temperature dependent measurement (Figure 3 and SI4). The CB spectrum, energy distribution curves (EDCs) at $\bar{\Gamma}$ and the normalized EDCs (with respect to the EDC at 105K) are plotted in Figs. 3a, b and c, respectively. The CB moves continuously towards $E_F$ and disappears upon increasing temperature, suggesting a Lifshitz transition in the bandstructure. The abrupt change appears at 100K-105K (Figs. 3a9-10) near $T_C$, where the CB disappears. Such an evolution is an intrinsic property of *n*-HgCr$_2$Se$_4$ since the CB would reappear below $E_F$ after cycling the temperature back to 20K (Fig. 3a12). Contrarily, negligible changes are found in the VB (SI4) upon temperature variation.

Above $Tc$, there seems to be no density of states near $E_F$ (Figs. 3a10-11) in contrast to the observed electron pocket near $E_F$ below $T_C$ (Figs. 3a1-8). Such an evolution of the Fermi surface could be explained by considering the impurity bands from Hg and Se vacancies. Our *ab-initio* calculations demonstrate that the impurity bands from the Se vacancies lie across the CB bottom in the FM state (Fig. 3d). When the CB moves upward in the PM state, the impurity bands [40] lie within the band gap and accommodates the electrons from the CB. In return the impurity bands pin the $E_F$, leading to the observed bandstructure above $T_C$ (see SI4). The electrons are transferred back to the CB from the impurity bands when the CB moves downwards to cross the impurities bands in the FM state.

**Discussion and conclusion**

We summarize our temperature dependent results in Fig. 4a. The disappearance of the CB across $Tc$ in $n$-HgCr$_2$Se$_4$ suggests the band shift is associated with the ferromagnetism, which splits the CB via exchange interaction. On the other hand, the VB is not significantly affected by the ferromagnetism, *most likely due to its orbital characters*. The VB contains mostly Se $p$ orbitals, which has small coulomb energy $U$ and thus much smaller band splitting than the CB with more Cr $d$ orbitals [16-17, 37] (See SI5 for a comparison of calculated bandstructure below and above $T_C$).

The observed Lifshitz transition across the FM-PM transition provides the electronic origin of the HMST as well as the CMR, as summarized in Fig. 4b. The transfer of electrons between localized impurities bands and the itinerant CB well accounts for the large variation in the resistivity in the HMST across $Tc$. Meanwhile, the external magnetic field aligns fluctuated spin moments near $T_C$ (inset in Fig. 4b) [25, 36-37]. The exchange splitting strength is thus tuned, leading to a magnetic field driven Lifshitz transition (Fig. 4b(ii)) and the HMST (Fig. 4c), thereby achieving the CMR (Fig. 4d). Such a mechanism from the electronic structure point of view may be much simplified (*e.g.*, neglecting the defects scattering and the magnetic polarons [27-28, 41-42]), however it captures the essential physics. With different $E_F$ positions due to different impurity levels, the HMST and CMR would change accordingly, also showing nice agreement with our interpretations (see discussions in SI6). The mechanism can also nicely explain the giant red shift of the optical absorption edge [35].

The absence of other complex correlated phenomena (*e.g.* phase separation [43-44], Jahn-Teller distortion [3, 20] and charged ordered state in manganites [44]; the extra incoherent band showing pseudogap in manganites [45] and Eu-chalcogenides [46], and the electron polaron coupled replica bands [47] in Eu-chalcogenides) in HgCr$_2$Se$_4$ offers us the excellent opportunity to identify the key ingredients for the HSMT and CMR in half-metals. Our findings could be further applied to other half-metal systems such as manganites and Eu-chalcogenides. Because the HSMT and CMR in these systems always appear in the same temperature regime near $T_C$ [3, 18-28], requiring additional turning knobs (*e.g.*, doping [3, 19-23, 48], strain [3, 19-20, 49], *et al.*) to achieve both ferromagnetism and ideal $E_F$ position.

Finally, we note that the observed band gap in FM state from ARPES and optical experiment contradicts with the predicted Chern semimetal state [16] in HgCr$_2$Se$_4$. This is probably because of the relative overestimate of the spin-orbital coupling strength and the amplitude of band inversion.

In summary, we reveal a Lifshitz transition in *n*-HgCr$_2$Se$_4$ that results from exchange band splitting at a proper $E_F$. The observed electronic transition plays the dominant role of the HMST and CMR in *n*-HgCr$_2$Se$_4$. Our findings highlight that achieving both strong ferromagnetism and ideal $E_F$ positions to fine-tune the electronic bandstructure are critical to the realization and application of the HMST and CMR among the spinel and other half-metal systems.

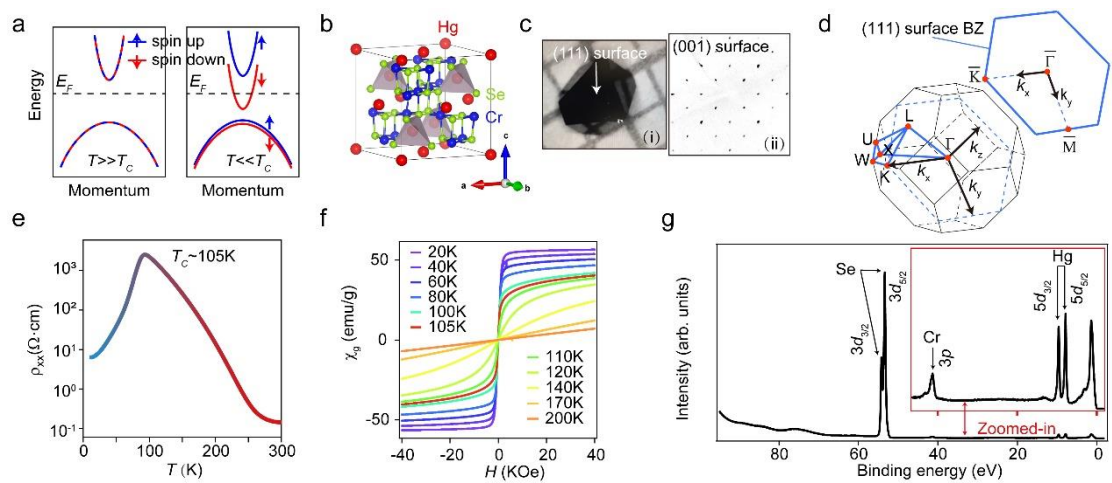

**Figure 1. Basic information and charaterization of *n*-HgCr$_2$Se$_4$.** (a) Schematic of band structure of *n*-HgCr$_2$Se$_4$ in the PM and FM states. (b) Crystal structure. (c) (i) Typical picture

of the sample. The natural facet is the (111) surface. (ii) Single crystal X-ray diffraction pattern. (d) Bulk BZ and (111) surface projected BZ. (e) Temperature dependent resistivity. (f) Temperature dependent magnetic susceptibility across $T_C$. (g) XPS spectrum with sharp Se $3d$, Cr $3p$ and Hg $5d$ core levels.

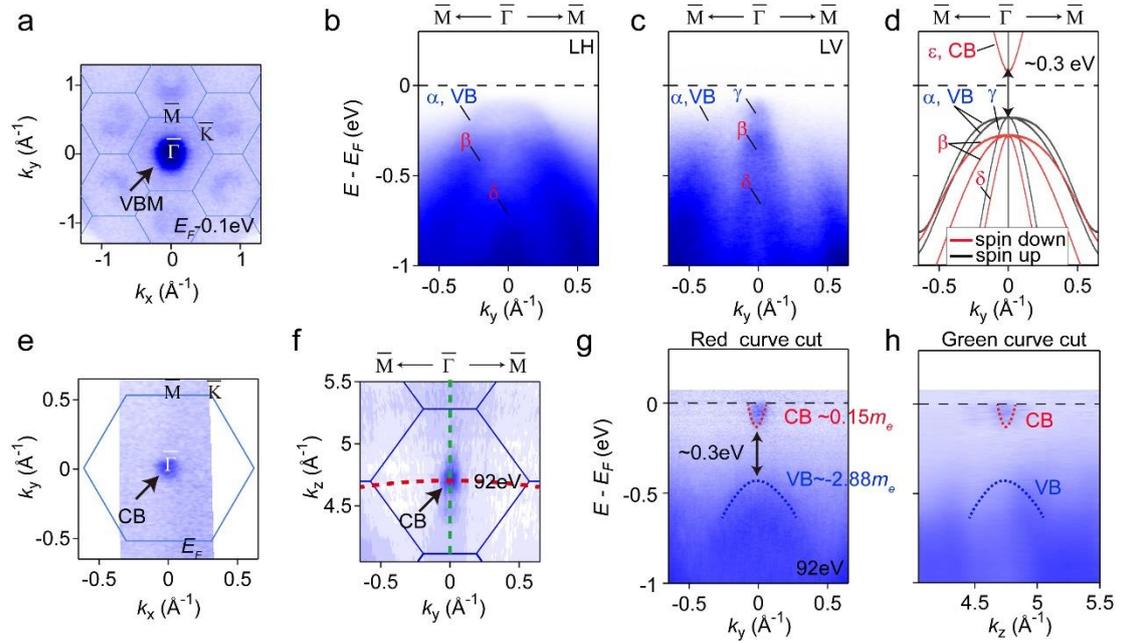

**Figure 2. Basic bandstructure of *n*-HgCr$_2$Se$_4$ on the (111) surface in the FM state.** (a) CEC which cuts through the VB maximum (VBM). (b-c) Band dispersions along $\bar{\Gamma}$-$\bar{M}$ measured by LH and LV photons. (d) Calculated spin-resolved bands along $\bar{\Gamma}$-$\bar{M}$. The bands ($\alpha$, $\beta$, $\gamma$, $\delta$, $\varepsilon$) are denoted accordingly. (e) CEC near $E_F$ after K deposition in the $k_x$-$k_y$ plane. (f) CEC near $E_F$ after K deposition in the $k_y$-$k_z$ plane. (g-h) Extracted bands along the red and green momentum curves in (f). The spectrum in (g-h) is normalized by dividing momentum-integrated EDC along the momentum direction. The measured temperature is ~20K.

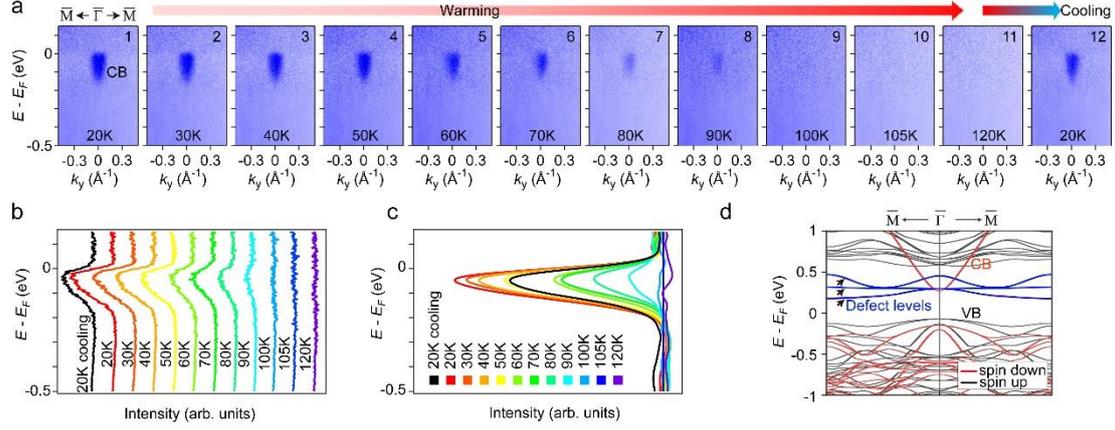

**Figure 3. Half-metal driven by ferromagnetism in *n*-HgCr$_2$Se$_4$.** (a) Evolution of bands along $\bar{\Gamma}$-$\bar{M}$ after K dosing at different temperatures. They are normalized in the same way as in Figure 2(g-h). (b) Extracted temperature dependent EDCs from (a) with a momentum integration of $\pm 0.025$ Å$^{-1}$. (c) Overlaid temperature dependent EDCs from (b). They are divided by the EDC at the temperature of 105K. (d) The spin-resolved band structure from the Hg$_{15}$Cr$_{32}$Se$_{63}$ supercell (Hg vacancy level is 1/16 and Se vacancy level is 1/64) which has defect levels from Hg and Se vacancies. The ones labeled by the blue color are mainly from the Se vacancies.

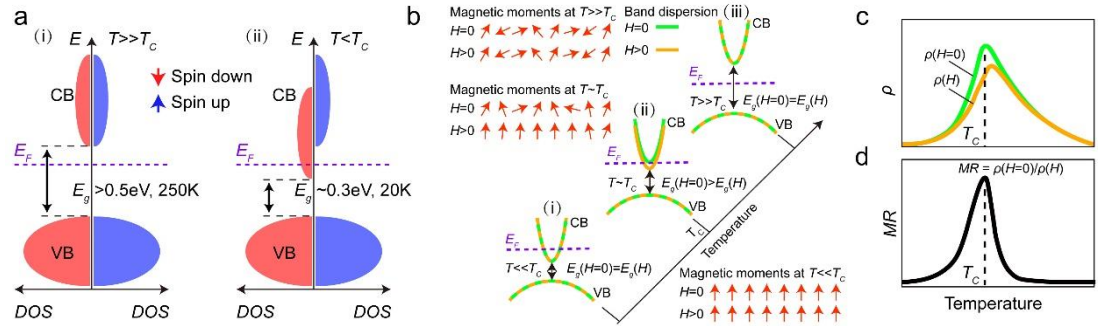

**Figure 4. Eletronic origin of the HMST and CMR in *n*-HgCr$_2$Se$_4$.** (a) Schematic of the electronic origin of the HMST. (i) $T \gg T_C$, the CB and VB have no spin related splitting. The band gap ($E_g$) is >0.5eV at 250K (SI4-5). (ii) $T \ll T_C$, the CB splits and crosses $E_F$ while the VB has minor changes (SI4-5). The band gap narrows to ~0.3eV at 20K. (b) Schematic of the electronic origin of the CMR. (i-iii) Temperature dependent bandstructure with and without magnetic field at $T \ll Tc$ (i), $T \sim Tc$ (ii) and $T \gg Tc$ (iii). Inset: Schematic of the alignment of the magnetic moments by magnetic field across $T_C$. (c) Schematic of the teperature dependent resistivity curve ($\rho - T$) of the HMST accroding to the electronic evolution in (b). (d)

Schematic of the teperature dependent magnetoresisitance showing CMR based on the $\rho - T$ curve in (c). Note that the $E_F$ position here is chosen according to the experimental observations in Figure 3.

**Acknowledgements:**

We thank B. Fu very much for the stimulating discussions. This work is supported by the grant from Chinese Academy of Science-Shanghai Science Research Center, Grant No: CAS-SSRC-YH-2015-01. Y.L.C. acknowledges the support from the Engineering and Physical Sciences Research Council Platform Grant (Grant No. EP/M020517/1) and Hefei Science Center Chinese Academy of Sciences (2015HSC-UE013). A.J.L. acknowledges the support from Scientific Support Group (SSG) Collaborative Postdoctoral Fellowship (P-1-06007) of the Department of Energy (DOE), Lawrence Berkeley National Laboratory (LBNL), and the Advanced Light Source (ALS). Z.K.L. acknowledges the support from the National Natural Science Foundation of China (11674229). ALS is supported by the US DOE, Office of Basic Energy Sciences, under contract No. DE-AC02-05CH11231.